# Tackling extreme urban heat: a machine learning approach to assess the impacts of climate change and the efficacy of climate adaptation strategies in urban microclimates

## Authors
Grant Buster[1], Jordan Cox[2], Brandon N. Benton[1], and Ryan N. King[3]

## Affiliations
1. Strategic Energy Analysis Center (SEAC), National Renewable Energy Laboratory (NREL)
2. Energy Security and Resilience Center (ESRC), National Renewable Energy Laboratory (NREL)
3. Computational Science Center (CSC), National Renewable Energy Laboratory (NREL)

Corresponding Author: Grant Buster (grant.buster@nrel.gov)

## Abstract
As urbanization and climate change progress, urban heat becomes a priority for climate adaptation efforts. High temperatures concentrated in urban heat can drive increased risk of heat-related death and illness as well as increased energy demand for cooling. However, estimating the effects of urban heat is an ongoing field of research typically burdened by an imprecise description of the built environment, significant computational cost, and a lack of high-resolution estimates of the impacts of climate change. Here, we present open-source, computationally efficient machine learning methods that can improve the accuracy of urban temperature estimates when compared to historical reanalysis data. These models are applied to residential buildings in Los Angeles, and we compare the energy benefits of heat mitigation strategies to the impacts of climate change. We find that cooling demand is likely to increase substantially through midcentury, but engineered high-albedo surfaces could lessen this increase by more than 50%. The corresponding increase in heating demand complicates this narrative, but total annual energy use from combined heating and cooling with electric heat pumps in the Los Angeles urban climate is shown to benefit from the engineered cooling strategies under both current and future climates.

## 1 Introduction

Urbanization is an impactful recent phenomenon in human history with substantial urban migration occurring in the last century. Currently, only 1% to 3% of the Earth's surface is considered urban[1]. However, although occupying a relatively small footprint, this small fraction of land contains about 55% of the world's population. This has increased from 30% in 1950 and is expected to increase to 68% by 2050[2]. This is an important factor in our shared climate future, as urban landscapes can contribute to local and regional warming[3] while affecting the per-capita utilization of various materials and energy resources[4].

Urban centers also create unique microclimates because of urban albedo, surface roughness, radiative trapping, and heat storage[5,6]. One well-known urban phenomenon is the urban heat island (UHI), in which the combination of built environmental factors causes the urban temperature (both surface and air) to be hotter than its rural counterpart[6–9]. This is not simply an inconvenience; urban extreme heat is a key driver of mortality risk and socioeconomic costs[10,11]. The population exposure to these risks in the United States is expected to increase in future years because of climate change and the size and spatial distribution of population[12]. Fortunately, adaptation strategies show promise in their ability to reduce UHI effects and could be used to combat this risk[13,14].

Modeling of UHI and urban heat mitigation strategies has historically been explored through a variety of methodologies that fall into several main categories: physics-based models with parametric inputs[5], reduced-order models based on the physics-based models[9], empirical models based on observations[8], and direct analysis of observations, including remotely sensed data fields[7]. Additional delineations can be drawn between research that characterizes heat climatology versus heat dynamics (e.g., annual average versus hourly UHI), varying spatial resolutions (e.g., citywide versus neighborhood impacts), impacts to surface temperatures versus air temperatures, the efficacy of climate adaptation strategies (e.g., albedo modification or enhanced vegetation), and the impacts of climate change. Previous studies that focus on modeling UHI from observations tend to focus primarily on remotely sensed land surface temperature (LST) data[7,15]. Conversely, UHI air temperature research is commonly performed using a low-spatial-resolution idealized physical model[5,16] or a high-resolution computationally expensive physical model[17].

We propose models for studying urban LST and near-surface air temperature at a high spatiotemporal resolution (nominal 500 meters [m] hourly) by learning physical phenomena directly from remotely sensed data fields, point ground measurements, and a large array of ancillary data describing the built environment. We validate the models using remotely sensed LST observations and near-surface air temperature measurements. We show the models can provide accurate estimates of urban heat, such as during the unprecedented 2021 Pacific Northwest heat wave, and they can estimate the effects of albedo manipulation as a heat mitigation strategy in Los Angeles. We integrate these capabilities with a physics-based residential building simulator to estimate the changes in energy use through midcentury as a result of climate change and the potential benefits from heat mitigation strategies. The models are built on previous generative machine learning super-resolution work called "Sup3rCC,"[18] and so we title these new models and methods "Sup3rUHI" (i.e., super-resolution for renewable resource data and urban heat islands).

## 2 Methods and Data

We implement a multistep modeling process, illustrated in Fig. 1, that learns what the spatial variability of LST is and how this high-resolution LST and the built environment drive local increases in near-surface air temperature and relative humidity. Although land and air temperatures are both important drivers of UHI, air temperature is typically more useful for studying public health impacts because it is the best indicator of conditions experienced by humans[19]. LST is learned from spatially continuous satellite observations from Moderate Resolution Imaging Spectroradiometer (MODIS),[20–22] whereas the local air temperature effects are learned from pointwise ground measurements from Meteorological Assimilation Data Ingest System (MADIS)[23] and Heat.gov[24,25]. This builds on prior work by Tran and Liou[26], Alvi et al.[27], and Shandas et al.[25]. We deviate from this previous work by leveraging machine learning-based super-resolution methods that can impute urban effects into low-resolution data fields from both historical and future weather and climate inputs at an hourly resolution for extreme heat wave dynamics.

Building on previous work by Buster et al.[18], the LST model is based on a fully convolutional generative adversarial network (GAN) using the sup3r software and is further described in Section 2.1. The near-surface air temperature and relative humidity model is based on a feedforward network with a custom layer that learns a distance-decay effect from a combination of raster and pointwise data, further described in Section 2.2.

A key contribution of this work is to integrate a large number of high-resolution satellite observations and urban settlement layers to the training and inference process, which helps the model output a spatially accurate heat climatology. All the datasets used in this work are listed in Table 1. The datasets used are free and publicly available but take considerable effort to process onto a common grid and format. As part of this work, we have released several years of preprocessed training data for the top 50 most populous cities in the contiguous United States. The data is in a common GIS-compliant format and can be used to build additional or similar heat resilience models. We have released cleaned air temperature and humidity observations from MADIS only for our two validation cities because of the considerable effort required to review and clean this data, but international measurement data is publicly available via MADIS web tools and can be processed for new cities in follow-on work as needed.

Note we discuss nominal grid resolutions in meters for convenience, but the data has true north-south and east-west spacing of 0.0083°, which results in true spacing of 460 m × 385 m for Los Angeles and 460 m × 310 m for Seattle. Different area- or distance-preserving geographic projections might result in more accurate models and might be explored in future revisions of the data.

For additional details on data and software, see Section 7.

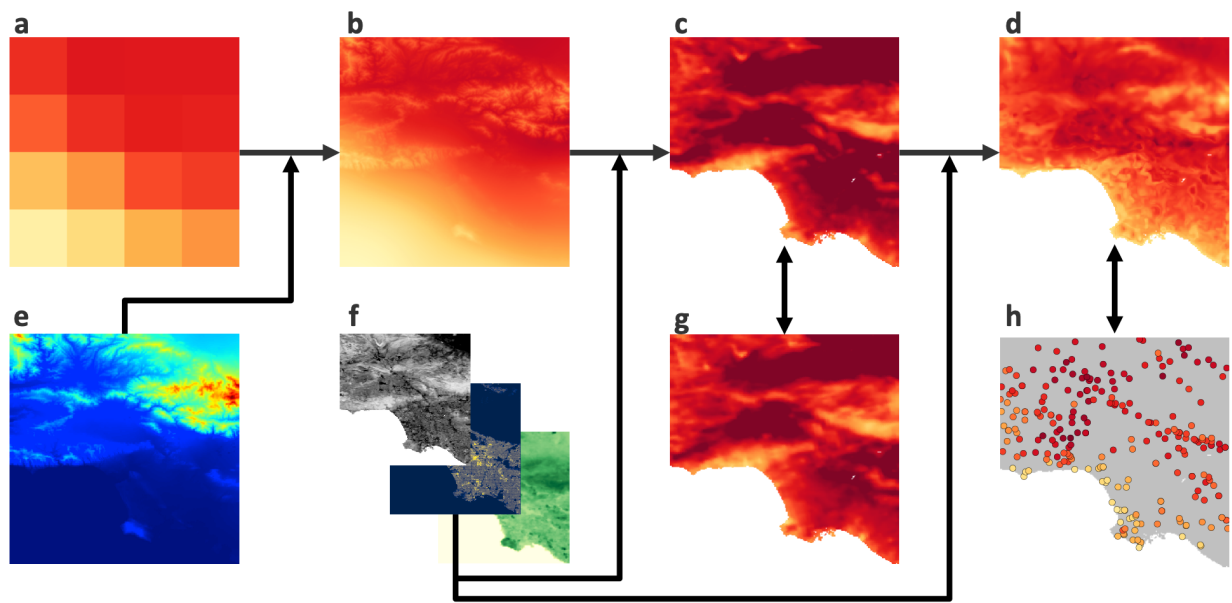

**Fig. 1. a-h,** UHI modeling process outline showing example data from Los Angeles. **a,** Low-resolution air temperature data from European Centre for Medium-Range Weather Forecasts Reanalysis 5 (ERA5). **b,** Interpolated ERA5 data using lapse rates. **c,** LST estimated using Sup3rUHI. **d,** Near-surface air temperature estimated using Sup3rUHI. **e,** Surface elevation data used as input to perform lapse rate interpolation. **f,** Ancillary data inputs such as surface albedo, built environment, and vegetation. **g,** Observed LST from MODIS satellites used for model training. **h,** Observed near-surface air temperature from the MADIS sensor network used for model training.

Table 1. Data sources used for urban heat island modeling.

| Variable | Units | Resolution | Source |
| --- | --- | --- | --- |
| Land surface temperature (LST) | °C | 1 km 12 hourly | MODIS[20–22] |
| Surface albedo bands 1-7 | Unitless | 500 m daily | MODIS[28] |
| Enhanced Vegetation Index (EVI) | Unitless | 1 km 16 day | MODIS[29] |
| Land cover | Unitless | 500 m yearly | MODIS[30] |
| Topography | m | 90 m | SRTM[a31] |
| Population | Number | 100 m | GHSL[b32] |
| Built volume | m$^3$ | 100 m | GHSL[b33] |
| Built height | m | 100 m | GHSL[b34] |
| Air temperature (reanalysis) | °C | 31 km hourly | ERA5[35] |
| Relative humidity (reanalysis) | % | 31 km hourly | ERA5[35] |
| Wind components | m/s | 31 km hourly | ERA5[35] |
| Sea surface temperature | °C | 31 km hourly | ERA5[35] |
| Air temperature (observed) | °C | Pointwise subhourly | MADIS[23] |

| Relative humidity (observed) | % | Pointwise subhourly | MADIS[23] |
| Air temperature (observed) | °C | Pointwise subhourly | Heat.gov[24,25] |
| Relative humidity (observed) | % | Pointwise subhourly | Heat.gov[24,25] |

[a] shuttle radar topography mission
[b] global human settlement layer

## 2.1 Land Surface Temperature Model

The LST model is based on the GAN architecture originally developed by Stengel et al.[36] and further extended in Buster et al.[18]. Similar to the models described in Buster et al.[18], each generator model is a deep, fully convolutional neural network[37] with 16 residual blocks[38] with skip connections[39,40]. The convolutional kernels are 3 × 3 × 3 and convolve across both spatial and temporal dimensions[18]. Each convolutional block is followed by a leaky rectified linear activation function with the negative slope coefficient set to 0.2[18]. Unlike previous work, the LST model developed here predicts high-resolution urban LST without enhancing the spatial or temporal dimensions in the network's forward pass and instead relies on interpolated input (e.g., with the lapse rates used by Buster et al.[18]) or known high-resolution urban layers (e.g., the 100-m population data from Schiavina et al.[32]).

The LST model is trained using the GAN approach[41] with two competing deep convolutional networks: one generator network, $G$, and one discriminator network, $D$. The LST generator network adds the urban effects to the input data fields $G: \chi_{input} \to \chi_{urban}$. In this case, $\chi_{input}$ is interpolated atmospheric conditions that do not represent the urban microclimate along with known high-resolution urban layers, and $\chi_{urban}$ is the output data that represents high-resolution urban LST. The discriminator network performs the classification task $D: \chi_{higher} \to [0, 1]$, where the output is the probability that a given observation is true (e.g., from the source training data by Zhang et al.[20]) and not synthetic (e.g., generated from the generator network). This training procedure can be expressed as a minmax optimization problem[41], as in equation (1):

$$\min_G \max_D \mathbb{E}\left[\log(D(y))\right] + \mathbb{E}\left[\log(1 - D(G(x)))\right] \qquad \textbf{equation (1)}$$

where $y \in \chi_{urban}$ is the observed high-resolution urban LST data and $x \in \chi_{input}$ is the input interpolated weather data without urban effects and known urban layers. In this case, the generator $G$ seeks to minimize equation (1), whereas the discriminator $D$ tries to maximize it. The loss function for the discriminator is then defined as in equation (2):

$$\mathcal{L}_D(x, y) = -\log(D(y)) - \log(1 - D(G(x))), \qquad \textbf{equation (2)}$$

where we seek to minimize $\mathcal{L}_D$. We define the generator loss function to include both an adversarial loss and a content loss, as in equation (3):

$$\mathcal{L}_G(x, y) = \mathcal{L}_{content}(x, y) + \alpha \mathcal{L}_{adverserial}(x, y) \qquad \textbf{equation (3)}$$

where $\alpha \in \mathbb{R}$ is a scaling term, $\mathcal{L}_{content}$ is the measure of the pointwise error of the true high-resolution data $y$ versus the synthetically generated high-resolution data $G(x)$, and $\mathcal{L}_{adverserial}$ is a binary cross-entropy loss defined as in equation (4):

$$\mathcal{L}_{adversarial}(x, y) = -\log(D(G(x))) \qquad \text{equation (4)}$$

The generator and discriminator networks are trained using the same pretraining and loss-balancing scheme described in Stengel et al.[36]: the generator is pretrained for a number of epochs with $\alpha = 0$. During the main training phase with $\alpha > 0$, the generator is trained only if the discriminator loss is less than 0.6, and the discriminator is trained only if its loss is more than 0.45.

We train the LST model on data from the top 50 most populous cities in the contiguous United States using data from years 2018 and 2019 and using 2020 and 2021 for cross validation. We find that the LST model performs best when trained on data from all cities and then fine-tuned on data from a single city. Therefore, we use two LST models in the final results: one for Los Angeles and one for Seattle.

## 2.2 Air Temperature and Relative Humidity Model

Next, we train a model that performs the function $F: \chi_{input} \to \chi_{t,rh}$ to learn the local air temperature and relative humidity $(\chi_{t,rh})$ from the same interpolated atmospheric inputs, known urban layers, and the urban LST $(\chi_{input})$. Because air temperature and humidity cannot be remotely sensed over a continuous spatial extent, we need to train this model using point observations that can be extended to all locations in the spatial extent. Without observations that are continuous in both space and time, we cannot train a convolutional generative adversarial network like the LST model. Instead, we focus on a simple feedforward network with the loss function, as shown in equation (5):

$$\min_f \ \mathcal{L}_{t,rh}[F(\chi_{input}), Y] \qquad \text{equation (5)}$$

which minimizes the empirical loss $\mathcal{L}_{t,rh}$ of the temperature and relative humidity model predictions $F(\chi_{input})$ versus the ground observations $Y$.

We train the air temperature and relative humidity model on years 2019 and 2020 and use 2021 for cross validation. We find that the model benefits from data representing as much spatial diversity as possible. Accordingly, we train a single model using data from both Los Angeles and Seattle. We hold out approximately 10% of the ground measurement stations from both cities for spatial cross validation.

Previous work has asserted that sparse point measurements are insufficient to properly resolve realistic spatial temperature gradients[15,42]. In practice, this means that training a model with only the local pixel inputs and pointwise measurements results in artificially large temperature gradients because the model cannot incorporate effects from neighboring pixels on local temperatures. Other studies have used averages over moving spatial windows to incorporate the effects of land use and land cover over an extended spatial footprint to predict temperatures at a point location[25]. We implement a similar approach using a custom layer that learns the standard deviation of a gaussian smoothing kernel. This layer acts like a common max or mean pooling layer[43] but with a gaussian kernel that learns a distance-decay effect. The layer has a kernel size $N \times N$ that reduces the spatial dimensions by $N - 1$. Here, we train on inputs with $N = 11$, so the model input spatial shape is 11 × 11 and the output is 1 × 1, which can be used with the ground measurements.

This approach resolves several issues at once: the layer provides a trainable spatial smoothing effect that creates realistic spatial gradients in urban temperatures, it prevents the memorization of urban features that convolutional layers can exhibit, and it can be executed in a computationally efficient convolutional operator for a large spatial domain.

In summary, although there are challenges in training on sparse pointwise observations, we show here that a properly designed machine learning model coupled with an extensive ground measurement network can produce realistic spatial gradients (as shown in Fig. 2) and accurate temperature estimates under spatial cross validation. Future work could focus on advancing the accuracy of the custom gaussian pooling layer with more diverse training data or experimenting with flexible resolution graph networks for interpolation and prediction in sparse point clouds[44].

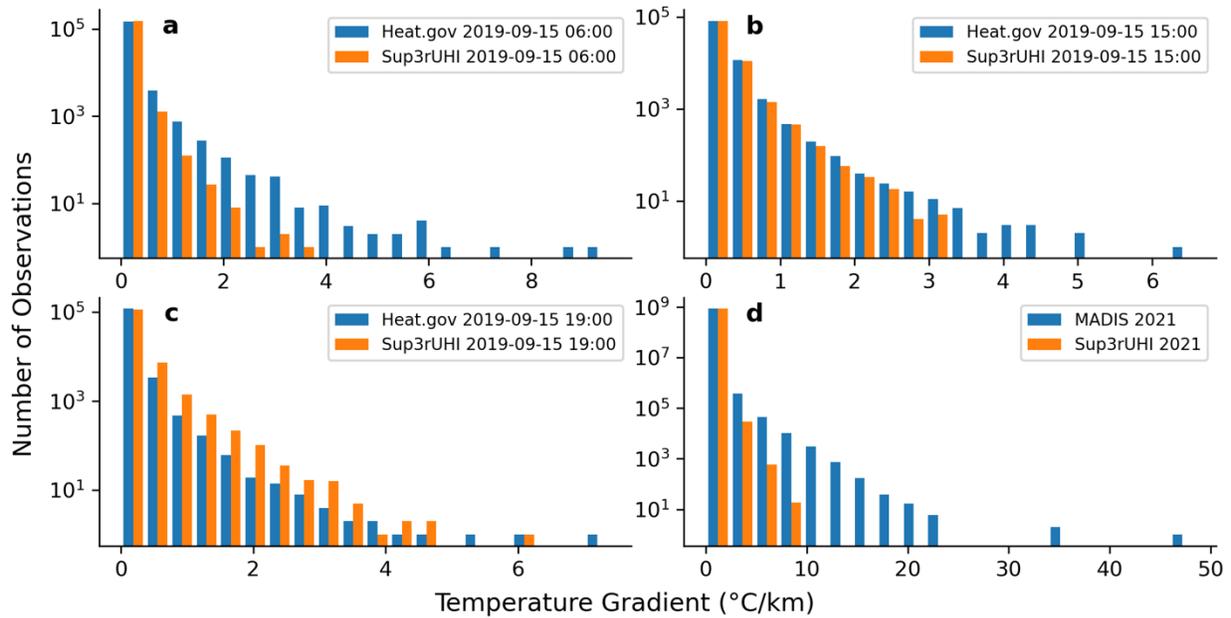

**Fig. 2. a-d, Histograms of modeled and observed temperature gradients in Los Angeles for three Heat.gov mobile traverse observations in 2019 at 0600 (a), 1500 (b), and 1900 local time (c) and for fixed MADIS ground measurement stations for all hours in 2021 (d). Temperature gradients are only calculated between measurements separated by more than 500 m.**

## 2.3 Residential Building Stock Model

To estimate the impacts of urban heat, climate change, and engineered heat interventions, we use a physics-based building stock energy model called ResStock v3.2.0[45], which represents prototypical housing stock across the entire United States.

To perform a focused regional study, we down select the residential building stock sampling configuration to create 100 prototypical buildings in Los Angeles County for the Public Use Microdata Area (PUMA) which represents North Central, Arleta, Pacoima, and San Fernando

Cities. Seven building simulations failed with invalid cool roof configurations, resulting in 93 viable building configurations studied. The PUMA studied here has approximately 375 housing units, so each sample in our simulation represents almost four real world housing units. Another four buildings in our sampled building stock were modeled as vacant and were excluded from the results presented in Section 3.4, since these buildings have different internal temperature than an occupied housing unit. Therefore, 89 occupied buildings are simulated with results presented in Section 3.4. The heating and cooling technologies represented in these buildings are described in Table 2 below.

**Table 2. Summary of HVAC technologies in the modeled building stock of 89 simulated occupied homes representing the PUMA for North Central, Arleta, Pacoima, and San Fernando Cities.**

| HVAC technology | Number of simulated homes with respective HVAC technology | Percent of total occupied building stock |
|---|---|---|
| Electric cooling | 58 | 65% |
| Gas heating | 53 | 60% |
| Electric cooling + gas heating | 40 | 45% |
| Heat pump cooling + heating | 6 | 7% |

We then create new weather files using the outputs of the temperature and relative humidity model alongside irradiance and wind data from Sup3rCC[18] based on the climate model EC-Earth3 for the SSP5-8.5 emissions scenario. The ResStock model then simulates the hourly residential building energy usage that would result from this climate-impacted weather timeseries data.

We use weather data from the pixel nearest to the Pacoima neighborhood because of the recent research done in this neighborhood on albedo modifications[46]. We build the weather files based on the ResStock typical meteorological year (TMY) file for Los Angeles County, holding the design conditions, typical/extreme periods, and other file headers constant while manipulating the hourly timeseries weather data. We run simulations for weather years 2015 to 2059, representing the current climate and one possible warming scenario.

We also simulate the building stock with and without a "Cool Roof" upgrade that maintains the roof material but modifies the roof color to white or cool colors for asphalt, metal, ceramic, concrete, and composite roof materials. This was configured using five standard ResStock upgrade options: "Roof Material|Asphalt Shingles, White or cool colors", "Roof Material|Metal, White", "Roof Material|Tile, Clay or Ceramic, White or Cool Colors", "Roof Material|Tile, Concrete, White or Cool Colors", and "Roof Material|Composition Shingles, White or cool colors".

Note that this experimental design is intended to isolate the impacts of climate change and heat adaptation measures on energy demand and indoor temperatures for the current residential building stock. This does not consider any updates to building heating, ventilating, and air-conditioning (HVAC) systems, new building construction in future years, or efficiency upgrades other than the rooftop albedo modifications that are discussed in Section 3.4.

# 3 Results

## 3.1 Model Validation

Maps of mean LST predicted by the Sup3rUHI model for Los Angeles and Seattle are presented in Fig. 3 and Fig. 4 alongside the mean error compared to gap-filled satellite observations from MODIS[20–22]. We present data only for summer months June, July, and August in these figures to focus on hot extremes, but the models perform similarly under all conditions, as shown quantitatively in Table 3.

The modeled mean instantaneous error tends to be below 2°C in Los Angeles with mean bias errors below 1°C. The Los Angeles model performs similarly in the pixels with the highest urban density and during the hottest observations. There is a negative bias in south Los Angeles in the Torrance and Long Beach areas but generally very low bias in the areas with the worst urban heat, such as downtown Los Angeles, Arcadia, San Fernando, and Panorama City.

In Seattle, we see mean instantaneous errors typically below 3°C. The Seattle model generally has a positive bias, including in the urban core; however, this is still generally below 1°C. Areas with the most intense urban heat, such as the area around the King County International Airport and Kent, tend to have very low bias.

Although the modeled LST values do exhibit error compared to satellite observations, the low bias error and local hot spots concentrated in urban areas give us confidence that this data is realistic and can be used in UHI assessments. However, although LST is a useful metric for UHI, near-surface air temperature is a much more important determinant of human health and safety.

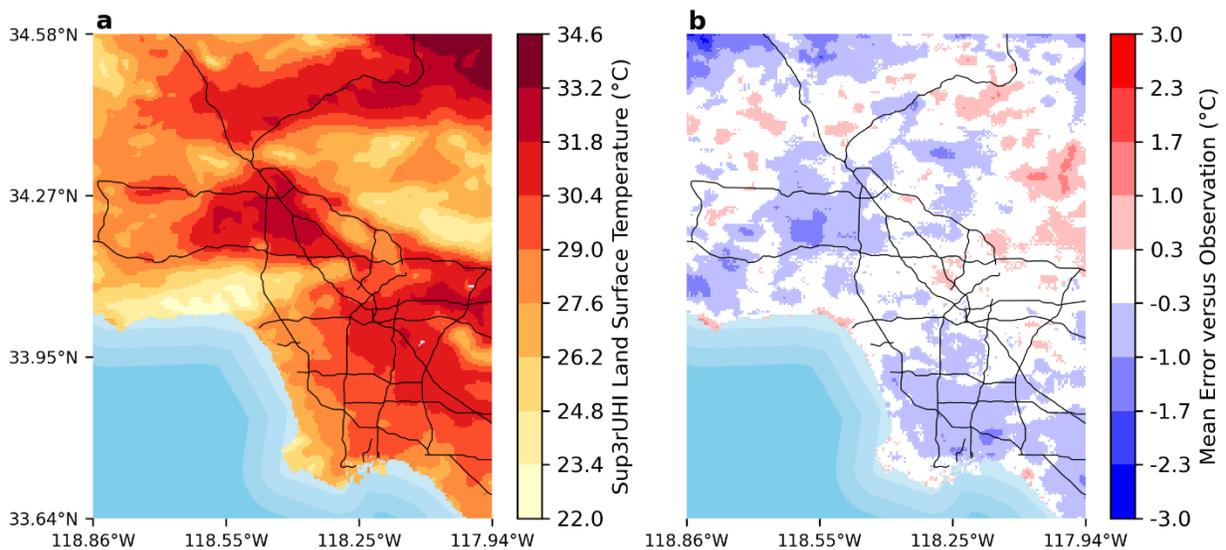

**Fig. 3. a, Mean predicted LST and b, mean error compared to satellite observations for the city of Los Angeles for summer months June, July, and August of 2020.**

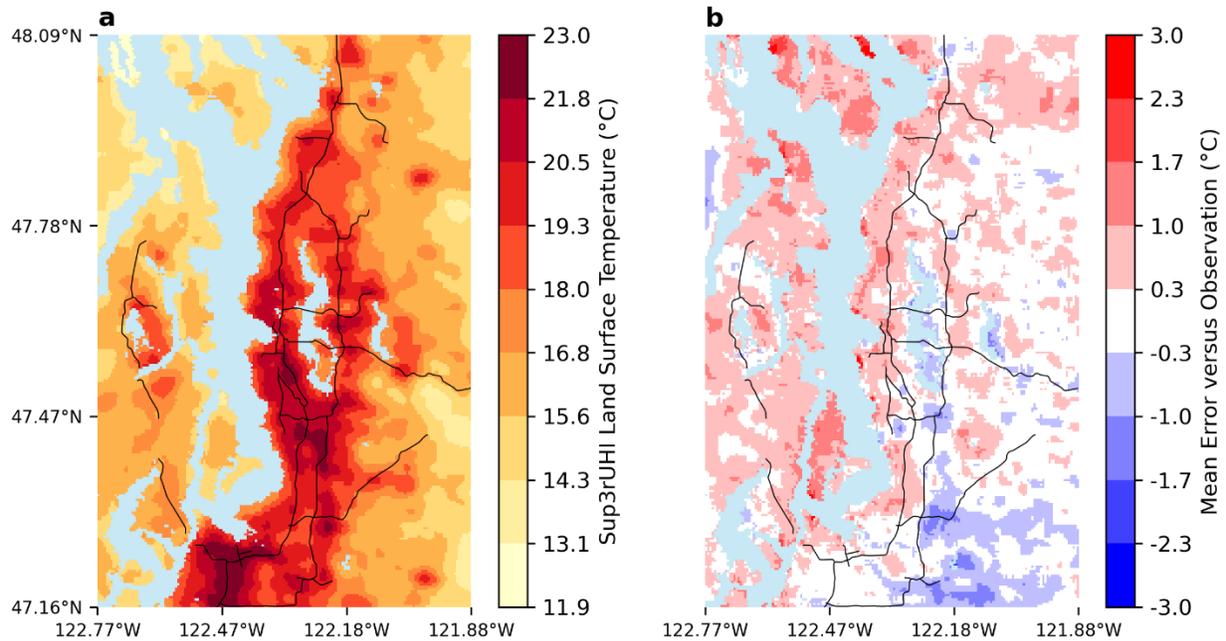

**Fig. 4. a, Mean predicted LST and b, mean error compared to satellite observations for the city of Seattle for summer months June, July, and August of 2020.**

**Table 3. Quantitative validation of LST output versus gap-filled MODIS data for all observations in 2020.**

| City | Condition | RMSE | MAE | MBE |
|---|---|---|---|---|
| Los Angeles | All data | 2.61 | 1.95 | -0.17 |
| Los Angeles | Top 10% built volume | 2.49 | 1.86 | 0.06 |
| Los Angeles | Top 10% hottest observations | 2.66 | 1.97 | -0.93 |
| Seattle | All data | 3.8 | 2.68 | 0.31 |
| Seattle | Top 10% built volume | 3.85 | 2.74 | 0.31 |
| Seattle | Top 10% hottest observations | 2.7 | 2.01 | -0.52 |

Maps of mean near-surface air temperature predicted by the Sup3rUHI model for Los Angeles and Seattle are presented in Fig. 5 and Fig. 6 alongside the mean error compared to ground observations from MADIS[23]. Similar to the LST figures, we present data only for summer months June, July, and August to focus on hot extremes, but the models perform similarly under all conditions, as shown quantitatively in Table 4.

In Los Angeles, we see that ERA5 frequently overestimates the air temperatures, especially in areas with high urban density. Sup3rUHI can reduce this error, with mean bias error (MBE) of -0.05°C overall and 0.11°C in the pixels with the greatest urban density. For all the results in Table 4, we see a general reduction in root-mean-square error (RMSE) and mean absolute error (MAE) with the biggest difference in the dense urban areas.

In Seattle, ERA5 does a much better job of estimating air temperatures, with lower bias and mean absolute errors than in Los Angeles under all conditions. Sup3rUHI is still able to improve upon the ERA5 data, however, especially in the bias during the hottest hours of the year. During these hot hours, ERA5 has a negative bias of -2.17°C, and Sup3rUHI can correct this to a bias of -0.72°C. This is discussed further in the next section with respect to the extreme 2021 Pacific Northwest heatwave.

Apparent in the temperature maps in Fig. 5 and Fig. 6 is the spatial variability in local climates even within a single city. Coastal regions in the Southeast of Los Angeles can be up to 9°C cooler on average than inland communities in the San Fernando Valley in the summer. Similarly, inland neighborhoods such as Georgetown in Seattle can exhibit temperatures up to 4°C hotter than islands in the Puget Sound. Choice of urban microclimate for weather inputs to building models can therefore have a significant impact on resulting estimates of energy demand. The data presented here can enable highly localized weather inputs as was used by Sandoval et al.[47] to study zip-code-level temperature equity.

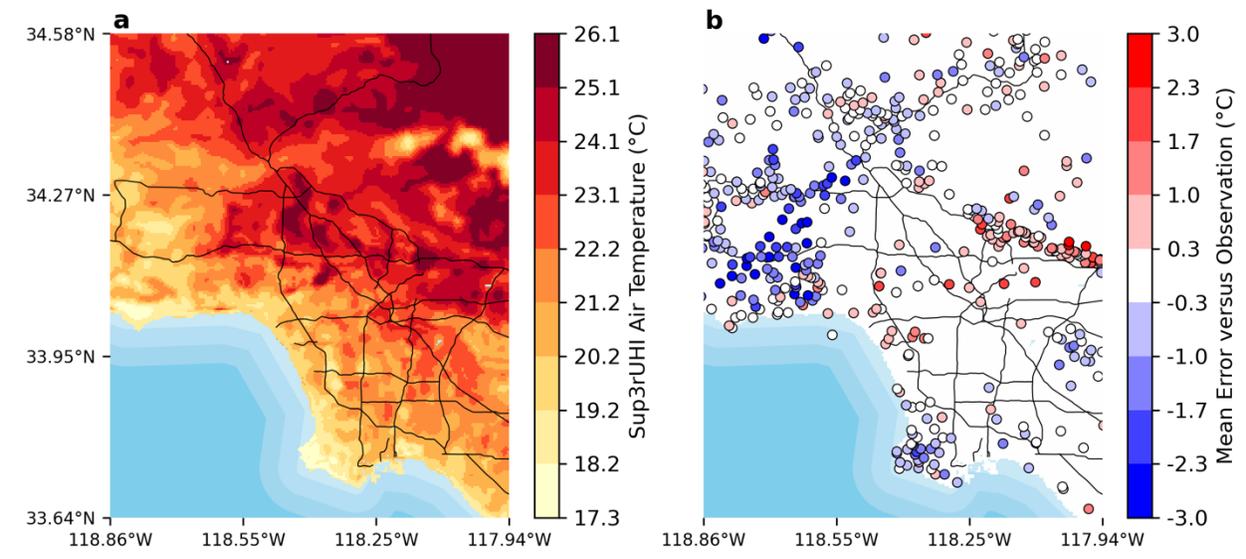

**Fig. 5. a, Mean predicted near-surface air temperature and b, mean error compared to ground measurement stations for the city of Los Angeles for summer months June, July, and August.**

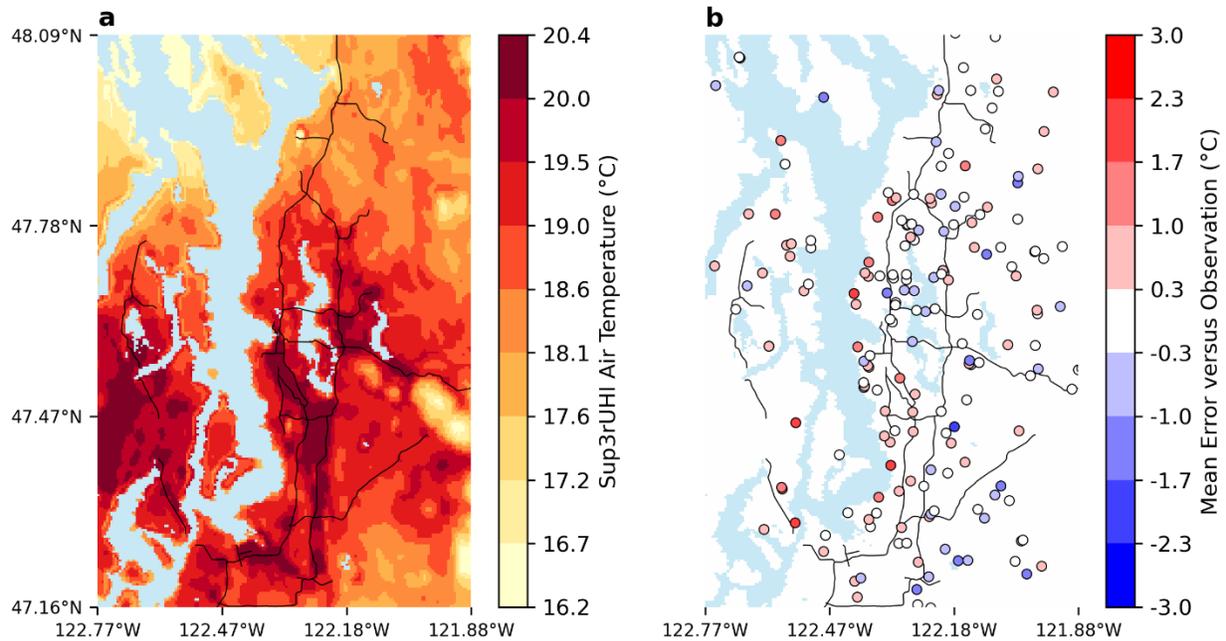

**Fig. 6. a,** Mean predicted near-surface air temperature and **b,** mean error compared to ground measurement stations for the city of Seattle for summer months June, July, and August.

**Table 4.** Quantitative validation of near-surface air temperature output versus ground observations from MADIS for all hours in 2021.

| City | Model | Condition | RMSE | MAE | MBE |
|---|---|---|---|---|---|
| Los Angeles | Sup3rUHI | All | 2.29 | 1.71 | -0.05 |
| Los Angeles | ERA5 | All | 3.01 | 2.30 | 0.65 |
| Los Angeles | Sup3rUHI | Spatial cross validation | 2.46 | 1.84 | 0.09 |
| Los Angeles | ERA5 | Spatial cross validation | 2.96 | 2.30 | 0.99 |
| Los Angeles | Sup3rUHI | Top 10% built volume | 1.86 | 1.40 | 0.11 |
| Los Angeles | ERA5 | Top 10% built volume | 2.69 | 2.07 | 0.95 |
| Los Angeles | Sup3rUHI | Top 10% hottest hours | 2.4 | 1.81 | -0.92 |
| Los Angeles | ERA5 | Top 10% hottest hours | 2.99 | 2.26 | -0.69 |
| Seattle | Sup3rUHI | All | 1.69 | 1.23 | 0.09 |
| Seattle | ERA5 | All | 1.83 | 1.33 | -0.01 |
| Seattle | Sup3rUHI | Spatial cross validation | 1.69 | 1.27 | 0.06 |
| Seattle | ERA5 | Spatial cross validation | 1.76 | 1.32 | 0.05 |
| Seattle | Sup3rUHI | Top 10% built volume | 1.44 | 1.08 | -0.14 |
| Seattle | ERA5 | Top 10% built volume | 1.55 | 1.14 | -0.24 |

| Seattle | Sup3rUHI | Top 10% hottest hours | 2.31 | 1.68 | -0.72 |
| Seattle | ERA5 | Top 10% hottest hours | 2.98 | 2.41 | -2.17 |

## 3.2 Pacific Northwest 2021 Heat Wave

In June 2021, the Pacific Northwest experienced a heatwave that many thought was impossible based on prior observations[48]. The event had catastrophic impacts, including hundreds of attributable deaths[49], and now serves as a crucial event for validation of urban heat models. Because we trained the air temperature model on years 2019 and 2020, the 2021 event significantly exceeds all Seattle heat events in the training data and serves as a cross-validation experiment.

Applying the Sup3rUHI models to the 2021 heat wave, we see improved performance when comparing the outputs to reanalysis data. Fig. 7 compares data from the Sup3rUHI model and ERA5 against four ground measurement stations. The ground measurement stations at Capitol Hill and the University of Washington are from NCEP MADIS and were selected as high-quality measurements from within the urban core while the two airport stations were taken from National Oceanic and Atmospheric Administration local climatological data. None of the four measurement stations were used in the training of the air temperature model.

Fig. 7 illustrates the excellent performance of the Sup3rUHI models when looking at daily peak temperatures in the urban core, for example, in Capitol Hill, University of Washington, and King County Airport. Both models underestimate daily peak temperatures at the Seattle Tacoma Airport. All models underestimate minimum nighttime temperatures, although we see some minor improvement with Sup3rUHI in Capitol Hill.

The performance of the Sup3rUHI model in the urban core during this unprecedented heat event supports the use of this model in studying similar unprecedented heat events driven by climate change.

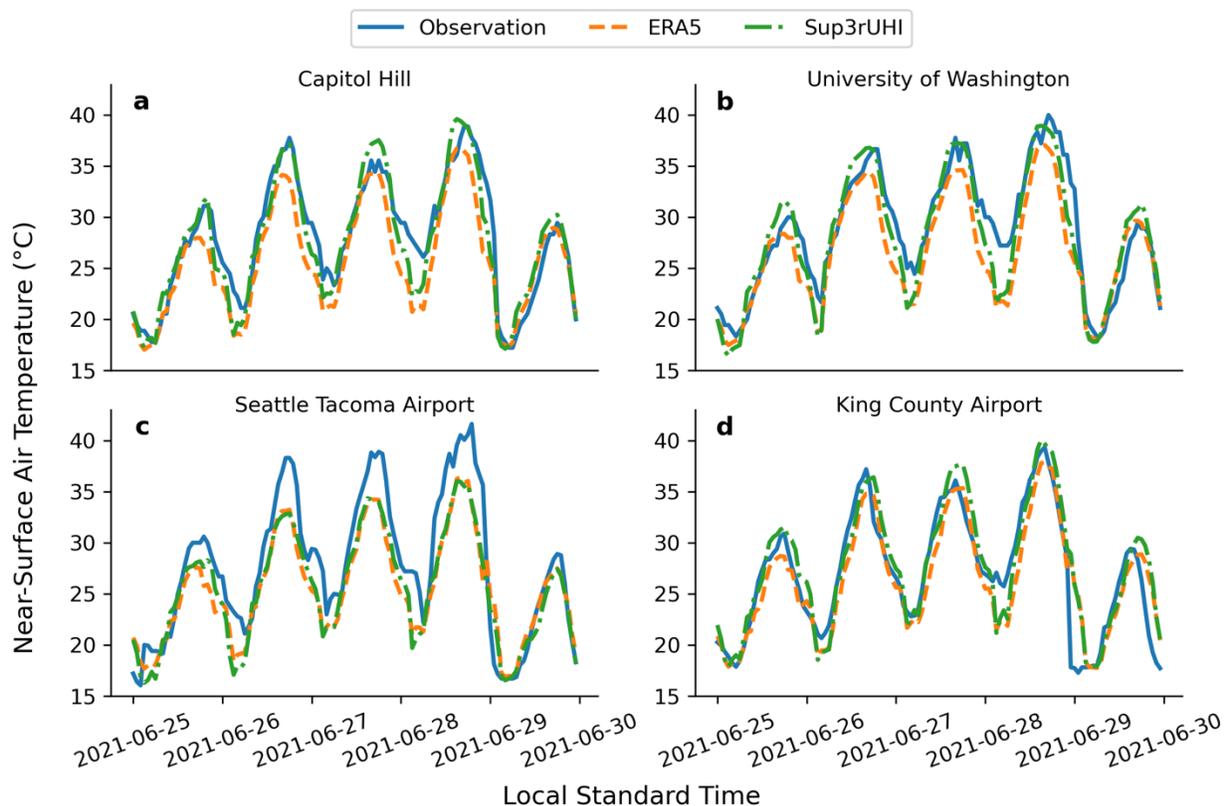

**Fig. 7. a-d, Timeseries of air temperatures observations, reanalysis data from ERA5, and modeled estimates from Sup3rUHI for the extreme 2021 Pacific Northwest heat wave at four ground measurement stations in Seattle: Capitol Hill (a), the University of Washington (b), the Seattle Tacoma Airport (c), and the King County Airport (d)**

## 3.3 Los Angeles Albedo Modification

Surface albedo modification has been identified as a potential solution to increasingly hot urban environments[50]. This climate change adaptation strategy has been studied using theoretical analysis[51], computer models[16,17,51,52], and more recently as pavement coatings in real-world neighborhood-scale experiments[46,50,53]. Air temperature reduction results of these experiments are presented below in Table 5, and it is obvious that the results vary significantly. Much of the variance can be attributed to methodological differences (e.g., to what extent the albedo was modified) and to the metric reported.

Generally, mesoscale meteorological models show average temperature reductions from 0.18 to 0.86°C[16,51], although these reductions can be as high as 1.1°C for single times[52]. One study using computational fluid dynamics models at very fine resolutions of 3 km showed differences up to 2.0°C, but these dramatic changes were observed only over the surface with modified albedo[17]. Real-world experiments have shown temperature reductions in the range of 0.2°C to 1.2°C using an adjacent neighborhood to calculate real-time and difference-of-difference estimates[46,50,53]. Temperature reductions are not reported in a consistent manner, and these estimates can vary by an order of magnitude depending on time of measurement and averaging across time and space, as shown in Table 5.

Using the Sup3rUHI model, we apply the road spectral albedo modifications reported by Taha[52] to the data inputs to both the LST and air temperature models and observe the resulting local changes in temperature. The modifications include average increases in dimensionless albedo of 0.04 from wavelengths 335-720nm and 0.23 from wavelengths 700-2500nm, which we apply to the seven MODIS albedo bands[28]. We focus on the Pacoima neighborhood because of the recent experiment there by Taha[52]. Based on arial imagery, we estimate that approximately 15% of the area in the Pacoima neighborhood was modified in the experiment by Taha[52], so we multiply the albedo changes by 0.15 before applying to our model inputs. We apply the changes to two neighboring grid cells totaling 0.35 km$^2$, which is approximately equivalent to the area modified by Taha[52]. We then run the Sup3rUHI models with all other inputs unmodified for the weather year 2021.

We find that the Sup3rUHI model estimates annual average air temperature reductions of 0.15 ± 0.13°C with a slightly greater reduction in the summer. This is comparable with the smaller reductions and seasonal effects we see in similar mesoscale models[16] and average observed real-time reductions in the real-world experiment by Taha[46]. We see instantaneous reductions up to 0.88°C, but these instances are very rare and we never see reductions up to the more extreme 2.8°C reported by Taha[46].

**Table 5. Comparison of reported air temperature reductions resulting from albedo modification experiments from this study and previous studies in the literature.**

| Study | Study Description | Air temperature reduction |
|---|---|---|
| This study | Machine learning model at 500 m with albedo modification in Pacoima, California | 0.15°C ± 0.13°C annual average, 0.18°C ± 0.09°C summer afternoon average |
| Taha[46] | Neighborhood-scale experiment and observation in Pacoima, California | 0.2°C -1.2°C mean afternoon differences, 1.4°C-2.8°C maximum afternoon differences |
| Ko et al.[53] | Neighborhood-scale experiment and observation in Covina, California | 0.2°C mid-day instantaneous |
| Middel et al.[50] | Neighborhood-scale experiment and observation in Sun Valley, California | 0.4°C -0.5°C afternoon instantaneous, negligible change after sunset |
| Taleghani et al.[17] | Neighborhood-scale computational fluid dynamics model at 3 m | 0.5°C -2.0°C afternoon instantaneous |
| Taha[52] | Mesoscale meteorological model at 1 km | 0.7°C -1.1°C daytime instantaneous |
| Mohegh et al.[16] | Mesoscale meteorological model at 4 km | 0.18°C -0.86°C annual average |
| Millstein and Levinson[51] | Mesoscale meteorological model at 0.44 km | 0.5°C ± 0.39°C mean daytime difference |
| Millstein and Levinson[51] | Theoretical hot plate analysis | 1.3°C |

## 3.4 Impacts of Climate Change and Heat Mitigation Strategies on Residential Energy Use

After showing the Sup3rUHI models can accurately estimate urban temperatures, unprecedented extreme heat waves, and the impacts of albedo modification on the urban microclimate (Sections 3.1, 3.2, and 3.3, respectively), we can now use the models to explore the impacts of urban heat, albedo modification, and climate change on residential building energy use. To do this, we combine temperature and relative humidity data from Sup3rUHI with solar irradiance and windspeed data from Sup3rCC[18] to create new weather files for input to the ResStock[54] physics-based residential building energy model.

We run four heat mitigation scenarios and one baseline scenario with no heat mitigation, further described in Table 6. The first two heat mitigation scenarios reduce outside ambient air temperatures with 15% and 30% albedo modification. These scenarios do not modify the building thermal envelope. Albedo modification of 15% represents modification of neighborhood roads and parking lots (e.g., the experiment carried out by Taha[46]), and 30% represents a hypothetical widespread effort that would additionally include additional surfaces such as rooftops, sidewalks, freeways, and driveways. The next heat mitigation scenario includes a "Cool Roof" upgrade that modifies the building thermal envelope in the ResStock simulation but does not modify the outside air temperatures. The last heat mitigation scenario is the most aggressive and includes "30% Albedo + Cool Roof" which reduces outdoor air temperatures via neighborhood albedo modification in addition to the Cool Roof installation that modifies the building thermal envelope. These four scenarios represent heat mitigation strategies for both the urban microclimate and the building thermal envelope.

Table 6. Summary of heat mitigation scenarios explored in this study.

| Scenario name | Neighborhood albedo modification (% of area) | Outside air temperature reduction in response to albedo modification (2015-2034 average) | Cool roof in ResStock building simulations |
|---|---|---|---|
| Baseline | 0% | 0.00°C | False |
| 15% Albedo | 15% | 0.16°C | False |
| 30% Albedo | 30% | 0.32°C | False |
| Cool Roof | 0% | 0.00°C | True |
| 30% Albedo + Cool Roof | 30% | 0.32°C | True |

All simulations are run with hourly timeseries for years 2015 to 2059 based on the climate model EC-Earth3 for the SSP5-8.5 emissions scenario[18]. Multiple possible climate change trajectories are available from the Sup3rCC dataset. We could explore the uncertainty in future climates, but we select a single possible climate trajectory to keep the analysis simple and to focus on the evaluation of urban heat mitigation strategies. Note that the SSP5-8.5 emissions scenario modeled using the EC-Earth3 climate model is one of the "hotter" climate change trajectories for the United States and particularly in the West[55]. Therefore, the mean changes through midcentury presented in Section 3.4 can be viewed as an upper bound on potential climate impacts.

All heat mitigation strategies lead to an immediate decrease in energy used for cooling in the current climate (Fig. 8a). The energy decreases range from 2% in the "15% Albedo" scenario to 10% in the "30% Albedo + Cool Roof" scenario. Based on the selected emissions scenario and climate model, climate change is estimated to increase cooling demand by 19% through midcentury, but this increase could be limited to 8% in the "30% Albedo + Cool Roof" scenario. Note, however, that these results are based on a fixed building stock in order to isolate the effects of climate change, and increased adoption and sizing of air conditioning systems under climate change[56] could lead to larger changes in cooling demand than are presented here. This uncertainty is discussed further in Section 4.

As climate change progresses, a warmer climate is likely to lead to an overall decrease in heating demand (Fig. 8b and Fig. 8c). However, the various heat mitigation strategies decrease temperatures during all times of year including winter months. The result is that, despite the warming climate, the most aggressive cooling strategy leads to an overall increase in heating demand from the current day baseline demand through approximately the mid-2030s.

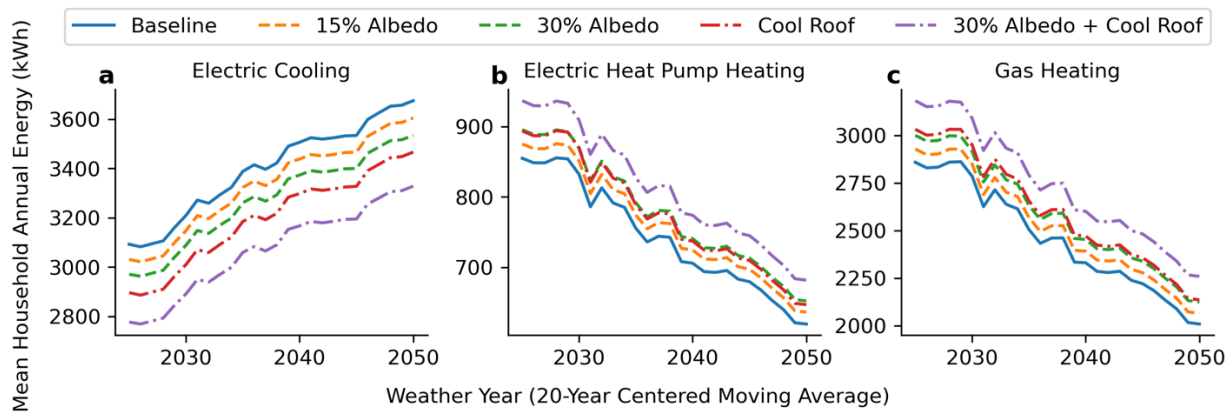

**Fig. 8. a-c, Changes in residential energy demand in response to climate change and several heat mitigation strategies for (a) homes with electric cooling (65% of building stock), (b) electric heat pump heating (7% of building stock), and (c) gas heating (60% of building stock). Note that each panel represents different subsets of the Los Angeles County residential building stock based on which buildings have the respective technologies.**

In the current climate, total energy demand from electric cooling and from gas heating are similar as are the respective changes in demand from the various heat mitigation strategies (Fig. 8a and Fig. 8c). This results in the heat mitigation strategies not having a significant effect on combined total energy demand from electric cooling and gas heating. Albedo modification could even lead to slight increases in overall combined total HVAC energy demand in the current climate (Fig. 9b and Fig. 9c). However, we find that climate change may reduce demand from gas heating more than it will increase demand from electric cooling. Therefore, the combined total energy demand from electric cooling and gas heating decreases through midcentury regardless of cooling strategy (Fig. 9b).

Heating by electric heat pumps changes the narrative considerably, although our modeled stock only had six homes with this technology. The overall heating energy demand from heat pumps is much less than from gas heating or from electric cooling (Fig. 8). The heat mitigation strategies increase heat pump heating demand but not nearly as much as gas heating demand. As a result, total heating and cooling demand from heat pumps increases through midcentury, and all heat mitigation strategies reduce this demand under both the current and future climate (Fig. 9a).

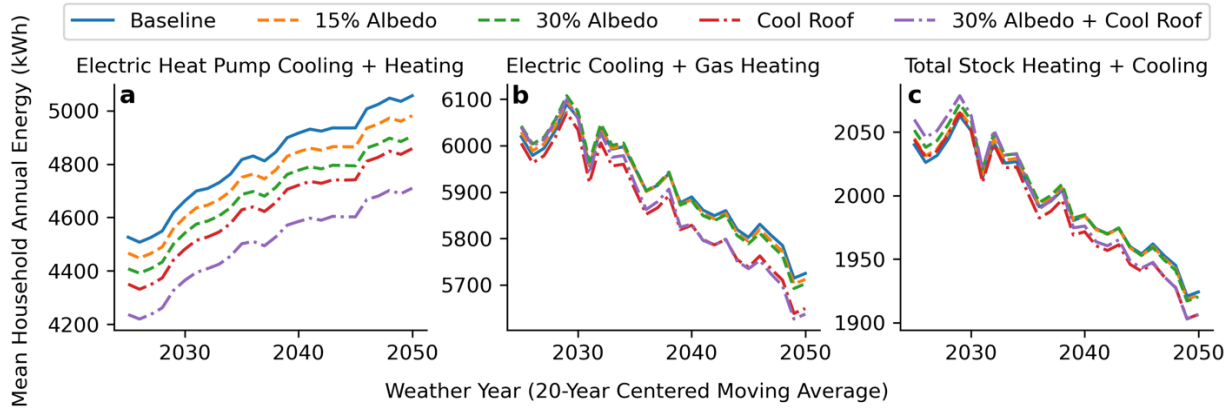

**Fig. 9. a-b, Changes in combined heating and cooling residential energy demand in response to climate change and several heat mitigation strategies for (a) combined heating and cooling with electric heat pumps (7% of building stock), (b) electric cooling combined with gas heating (45% of building stock), and (c) combined electric and gas energy demand for the entire building stock (100% of building stock). Note that each panel represents different subsets of the Los Angeles County residential building stock based on which buildings have the respective technologies.**

Based on these results, heat pumps and albedo-based heat mitigation strategies appear to be an effective combination for reduction in overall energy usage, even when considering the effects of climate change through midcentury in the Los Angeles urban climate. The heat mitigation strategies are also effective for reducing summer electricity demand, although they do cause a corresponding increase in winter heating demand. The dual nature of this effect is such that if albedo modification was implemented today in a Los Angeles neighborhood with electric air conditioning and natural gas heating, we may see a minor increase in total annual energy use. As energy use transitions to more cooling and less heating in a warming climate, these albedo-based heat mitigation strategies appear to benefit energy conservation even with inefficient heating technologies.

However, considering the potential for some regional power systems to develop winter peak loads under future electrification[57], widespread adoption of heat mitigation strategies in these climates could actually result in colder winter temperatures and increased heating demand. Holistic analysis like the one performed here should be carried out for different regional climates and subsequently extended to larger-scale impacts on the grid to understand the benefits and risks of such strategies under climate change and electrification.

# 4 Discussion

Here, we have presented a comparison of the impacts of climate change versus the efficacy of heat mitigation strategies and the effects of both on the urban microclimate and residential heating and cooling energy demand. Prior work has studied one or more of these aspects, and to our knowledge this is the first study that has integrated all these components into a single analysis. To our knowledge, this has not been done before because of the considerable technical challenges and computational requirements in the spatiotemporal downscaling of future climate projections, modeling of the urban microclimate, and estimation of heat mitigation strategies.

The novel machine learning methods developed here appear to be a viable option for addressing these challenges in a highly computationally efficient manner. Further, these methods can be adapted to new cities by simply downloading additional publicly available data and training new models. The total human labor burden required to extend these models and perform additional analyses is therefore expected to be much less than the labor required to build and execute a representation of a new city in a physics-based model.

The questions that can be answered by these methods are wide ranging and could be of great utility to city and urban planners. One could ask: What would be the climate benefits of installing high-albedo pavement in all public parking lots? What neighborhoods are most at risk in future heat waves? And where should community cooling centers or air conditioning retrofits be prioritized? Although we did not validate the model's representation of vegetation here, one could also likely ask of the model: How much do vegetation requirements in construction regulations affect local temperatures?

Previously, these questions were likely expensive to answer, requiring commissioned studies that could take onerous amounts of time. Our society has likely implemented cooling strategies based on intuitive knowledge alone, risking unexpected effects such as the resulting increased heating demand shown in Section 3.4. The computational efficiency of the methods developed here should improve the interactive interrogation of these questions. To illustrate, running the Sup3rUHI models for one weather year at a 500-m resolution at an hourly frequency takes 17 minutes on a dual-socket Intel Xeon Sapphire Rapids 52-core processor (104 cores total) with 256 gigabytes of memory. This includes the time to read and format input data; simply running the LST and air temperature models only takes 6 minutes. Simulating a single neighborhood such as in the Pacomia albedo study in Sections 3.3 and 3.4 is much faster, taking only 3 minutes per weather year on the same hardware, including the time to read and format input data. This enables near-real-time interactive data-driven analysis of adaptation strategies in collaboration with city planners.

There are, however, many uncertainties remaining in this work that are not quantified here. For example, there is considerable uncertainty in how the impacted building stock will change in the future, in the estimates of some high localized heat islands in the Sup3rUHI outputs, in the data we use for model training, and in the magnitude of cooling effect from albedo modification. We do not attempt to quantify these uncertainties but discuss them below.

It is important to consider the building results in Section 3.4 as only the isolated impacts of climate change and hotter weather on a fixed building stock. Our methods do not consider improvements to cooling and heating technology or changes to the building stock via renovation, demolition, or new construction, including future changes to California building codes. For example, increased temperatures due to climate change could increase the adoption and sizing of air conditioning sytems[56]. In our results, the fixed cooling systems that were sized for the current climate are observed to be undersized by midcentury leading to unmet hours. Future work could use these methods to estimate local climate impacts alongside electrification, energy efficiency, and equity strategies[47] to develop a more holistic understanding of climate impacts and heat mitigation on cooling loads. Furthermore, our results do not consider the impact of policy changes which influence the buildings stock such as California or federal appliance standards which could impact the efficiency of equipment into the future.

While the spatiotemporal outputs from the Sup3rUHI models generally make intuitive sense and perform well in the quantitative validation, there are still some locations that have unintuitive temperature estimates. For example, there are locations in the West Hollywood Hills and in Panorama City in Los Angeles that are several degrees hotter than their surrounding areas. Both neighborhoods have more vegetation than the surrounding neighborhoods, casting doubt on whether they are actually heat islands. However, the West Hollywood Hills have much lower visible light albedo and dramatically more built volume than surrounding neighborhoods, and Panorama City has an order of magnitude greater population density than its surroundings. Because Sup3rUHI is a data-driven model, over reliance on these features is possible, and it is difficult to know the extent of local temperature bias without additional measurement stations. Although Sup3rUHI generally agrees with other public UHI estimates such as from Heat.gov[24,25] and Climate Central[8,58], similar unintuitive temperature estimates can be found in all data-driven UHI estimates. For example, Sup3rUHI, Heat.gov, and Climate Central all agree that the temperature in Downtown Los Angeles is on average approximately 1-2°C hotter than the surrounding areas, but Heat.gov and Climate Central both estimate Santa Monica to be a strong local heat island despite its location on the coast. Future work could focus on increased and improved measurement data to build confidence in more diverse local climates.

The measurement data we use in this work was taken from many sources, as documented in Table 1. Each data source has its own respective uncertainty that is difficult or sometimes impossible to incorporate in this effort. For example, the MADIS measurement data for near-surface air temperature is a quality-controlled dataset that we additionally cleaned of obviously erroneous values (e.g., temperatures exceeding 60°C in Los Angeles). However, when calculating observed spatial temperature gradients, we still found temperature differences exceeding 8°C between stations separated by 40 m, which seems unrealistically large, especially considering that gradients from the National Integrated Heat Health Information System's Heat.gov measurement campaign did not exceed 10°C/km. It is possible that the MADIS measurement stations are not well calibrated or that neighboring sensors are placed near acute heat sources and sinks. When evaluating temperature gradients in Fig. 2, we only present measurements separated by more than 500 m, which reduces this effect and corresponds to the Sup3rUHI output resolution. Nevertheless, future work should focus on improving the training data.

Albedo modification is estimated to have a wide range of effects on the local microclimate from study to study, as shown in Table 5. This is complicated by the fact not all studies use the same methods and magnitudes for albedo modification and not all studies report the same metrics to measure the cooling effect. Sup3rUHI estimates the effect to be on the lower end of possible responses, but it is nearly impossible to evaluate the accuracy of this because of the difficulty in getting precise ground-truth measurements. Although several experiments have attempted to measure the real-world effects of albedo modification, establishing a perfect control to an experiment in the real world is difficult because of the natural spatiotemporal variability of weather. More advanced machine learning architectures may be able to better estimate the uncertainty in this effect, and higher-resolution data may be able to increase the precision with respect to the affected surfaces.

We also discovered an interesting challenge related to the development of data-driven models to estimate the effect of albedo modification. As we developed initial experiments using broadband albedo data, the trained models asserted a positive relationship between albedo and air temperature. As we looked at urban heat island maps alongside satellite imagery, this is intuitive from a data perspective—hot urban centers such as downtown Los Angeles tend to have higher albedo than neighboring forests that are cooler but also appear darker. Training on spectral albedo data from the MODIS satellites appears to alleviate this problem. Using spectral albedo data is also a logical approach based on the spectral signature of the cool pavement coating measured by Taha[46]. Still, though, in the experiment presented in Section 3.3, we do not see a significant response in mean LST to the albedo modification. It is possible that the observed LST data is too strongly correlated with the reflective built environment to learn the more subtle relationship between albedo and radiative balance. Future work might explore this further and could train on higher-resolution data that could resolve individual streets, parking lots, and other dark hot surfaces with additional precision.

# 5 Conclusion

In this study, we introduce machine learning methods to introduce high-resolution UHI effects into low-resolution historical reanalysis and future climate model datasets that do not represent urban heat phenomena explicitly. Models trained to estimate the UHI in Los Angeles and Seattle are made publicly available along with open-source software and additional training data for the 50 most populous cities in the contiguous United States, enabling the extension of this work to additional urban environments.

We demonstrate that these methods can be used to evaluate the impacts of climate change and the effectiveness of heat mitigation strategies within the context of high-resolution urban microclimate modeling. By integrating these components into a single analysis, the developed methods offer a computationally efficient and adaptable solution for addressing the challenges associated with spatiotemporal downscaling, urban microclimate modeling, and heat mitigation estimation. This enables city and urban planners to address a wide range of questions and prioritize interventions in a timely manner.

However, uncertainties remain in the measurement data used for model training and validation, as well as the effect of engineered heat mitigation strategies on the local microclimate. Future work should focus on improving the quality and quantity of the training data, joining this research with dynamic building stock and future electrification scenarios, exploring additional data that further describe the relationship between surface albedo and radiative balance, and potentially using higher-spatial-resolution training data to enhance the precision of the models.

Overall, this work has the potential to contribute to the development of more resilient and sustainable urban environments in the face of climate change. The open-source models, software, and data should enable researchers and urban planners to better understand urban heat and the benefits and drawbacks of possible heat mitigation strategies in their respective cities.

# 6 Acknowledgements

This work would not have been possible without significant help and review from Janet Reyna, Tony Fontanini, and Katelyn Stenger on the ResStock configuration and analysis. The authors would also like to thank Meghan Mooney and Dan Bilello for their thoughtful reviews of the initial draft, and Jennifer Korte for her invaluable help editing the manuscript.

This work was authored by the National Renewable Energy Laboratory, operated by Alliance for Sustainable Energy, LLC, for the U.S. Department of Energy (DOE) under Contract No. DE-AC36-08GO28308. Funding provided by the Laboratory Directed Research and Development (LDRD) program at the National Renewable Energy Laboratory. The research was performed using computational resources sponsored by the DOE Office of Energy Efficiency and Renewable Energy and located at the National Renewable Energy Laboratory. The views expressed in the article do not necessarily represent the views of the DOE or the U.S. Government. The U.S. Government retains and the publisher, by accepting the article for publication, acknowledges that the U.S. Government retains a nonexclusive, paid-up, irrevocable, worldwide license to publish or reproduce the published form of this work, or allow others to do so, for U.S. Government purposes.

# 7 Data and Software Availability

The input and output data presented here is publicly available via the Open Energy Data Initiative Submission #6220 (https://data.openei.org/submissions/6220)[59]. The associated code is based on the sup3r software package[60]. UHI-specific code is undergoing internal review and will be made open-source upon publication of this manuscript.

# 8 List of Acronyms

| | |
|---|---|
| ERA5 | European Centre for Medium-Range Weather Forecasts Reanalysis 5 |
| GAN | generative adversarial network |
| GHSL | global human settlement layer |
| km | kilometer |
| LST | land surface temperature |
| m | meter |
| MADIS | Meteorological Assimilation Data Ingest System |
| MODIS | Moderate Resolution Imaging Spectroradiometer |
| MAE | mean absolute error |
| MBE | mean bias error |
| m/s | meters per second |
| NREL | National Renewable Energy Laboratory |
| PUMA | Public Use Microdata Area |
| RMSE | root-mean-square error |
| SRTM | shuttle radar topography mission |
| Sup3rUHI | super-resolution for renewable resource data and urban heat islands |
| TMY | typical meteorological year |
| UHI | urban heat island |